Alexander Lipton

**(1) Please explain your education and work experience in Physics.**
I was born in Moscow, the capital of the Soviet Union, to a family of lawyers. Among the extended family, we proudly count several distinguished physicists and mathematicians, who inspired me from a very young age. After passing entrance exams, I started my studies at the fabled Mathematical High School # 2 for mathematically gifted children and graduated from it, *Summa Cum Laude*.

Upon graduation from high school, I was fortunate to be accepted to the Department of Mechanics and Mathematics of Moscow State University. At the time, it was the best department of mathematics in the whole Soviet Union, and, arguably, in the world. (From what I hear, it is still excellent.) I was deeply impressed and motivated by several of my teachers, such as the legendary Profs. A. N. Kolmogorov, I. M. Gelfand, V. I. Arnold, Ya. G. Sinai, A. G. Kostuchenko, A. T. Fomenko, to mention but a few. All of them were remarkable both as researchers and teachers. After receiving my MSc, *Summa Cum Laude*, I accepted the position of research scientist at the Soviet Academy of Sciences.

At the same time, I started to work on my Ph.D. dissertation at Moscow State University. My thesis was (or seemed to be) somewhat abstract, the title being "Spectral Properties of Degenerate Systems of Differential Equations." However, many years later, some of the methods I developed in my thesis came very handily when I studied algorithmic trading, see Lipton *et al.* (2014). This fact is a neat illustration of the profound interconnectedness of mathematical, physical, and financial phenomena. At the Academy of Sciences, I was captivated with and rigorously studied oscillations of the Earth's magnetosphere. For my work on this topic, I received "The Best Young Geophysicist Award" by the Academy. The work on the magnetosphere resulted in my life-long fascination with magnetohydrodynamics. While still in Russia, I wrote a book on the subject with an emphasis on thermonuclear fusion, which is still in print today; see Lifschitz (1989). One of the motivations for writing this book was a natural desire to put my thoughts on the subject in order. The other was an aspiration to learn English since it became clear to me that there was no future for my family and me in the Soviet Union.

At the end of 1988, I left the country of my birth for the US. As my first destination was Boston, I went to MIT to talk to some of the academics whom I knew by reputation. Professor Bruno Coppi, who read my book, offered me a research position in his group, which I gladly accepted. At the same time, I started to look for tenure track jobs, and, after a postdoc position at the University of Massachusetts, and my fair share of rejections, I landed a job at the University of Illinois, where I spent about seven years, long enough to get a full tenured professorship. At the same time, I worked as a consultant at the Los Alamos National Laboratory, where I continued research on fluids and plasmas.

At the University of Illinois, I got fascinated with fluid dynamics and astrophysics and wrote several papers on the subject, which were very well received by the scientific community, see

Lifschitz and Hameiri (1992), Lebovitz and Lifschitz (1996). Surprisingly or not, subsequently, I was able to use some of the ideas developed in these papers in mathematical finance, specifically for studying stochastic volatility models, see Lipton and Sepp (2008).

In 1996, I started to develop an interest in financial mathematics. In 1997, I took a leave of absence from the University of Illinois and moved to New York (for the actual reasons, see my answer to the next question). I started as a quant for Bankers Trust, the leading derivatives trading investment bank, which was acquired by Deutsche Bank in 1999. I was lucky to hit the ground running and began to work with several brilliant colleagues on many exciting topics of the day, including passport options, which are quantitatively different from calls, puts, and other options known for at least two hundred years. We were the first to publish a definite paper on the subject, Hyer *et al.* (1997). I continued to work on this topic for a couple of years and eventually proved that the value of a passport option is half the value of a lookback put, Lipton (1999). For my work in this direction, I received the very first "Quant of the Year Award" by Risk Magazine in 2000.

While at Bankers Trust, I became interested in foreign exchange and started actively working on it. In 2001, I had enough results under my belt to write a book on the subject, Lipton (2001), which continues to be the standard reference in foreign exchange derivatives to this day. After Deutsche Bank bought Bankers Trust, I stayed at DB for a couple of years but eventually moved on to Credit Suisse. Shortly afterward, I joined Citadel Investment Group in Chicago, where I became a managing director and head of capital structure quantitative research.

From Citadel in Chicago, my career trajectory took me to Merrill Lynch in London, where I initially headed the credit analytics group. In 2008, at the height of the Global Financial Crisis, Merrill Lynch was bought by Bank of America. I stayed with the Bank of America and became the co-head of the Global Quantitative Group. While in London, I kept my connections with academia as an honorary professor, first at the Imperial College, and then at Oxford, supervising several Ph.D. students, including Ioana Savescu, currently an MD at BofA, and Vadim Kaushansky, now at Citadel.

After spending 10 years with BofA, I left it in 2016 to start developing my ideas about banking and financial ecosystem in earnest. Advancing these ideas made me an expert on blockchains, digital currencies, including central bank digital currencies and stable coins; I summarized some of my thoughts on the subject in Lipton (2018).

In 2018, I co-founded, with S. Karkal, A. Angelovska-Wilson, and I. Hines, a new company called Sila, which develops new financial system based on the stable coin concept and digital wallet & payments. Silamoney aims to transform the existing banking and payment system in its entirety and make it more accessible, rational, and efficient.

I also went back to my academic roots and became a Connection Science Fellow at MIT, and, more recently, a Visiting Professor and Dean's Fellow at the Hebrew University of Jerusalem.

My varied career in mathematics, physics, and finance taught me that universal mathematical constructs and the way of thinking rooted in physics permeate mathematics, physics, and finance, and make them virtually the same. There is one caveat, though. One needs to understand how to formulate the problem. Otherwise, one can find himself/ herself in a vicious circle when people address the problem, they <u>can</u> solve, rather than the one they <u>need</u> to solve.

**(2) What "calculations" prompted you to move from physics to finance?**
The "calculations" were rather straightforward. In 1997 my wife graduated from the University of Chicago with two degrees - a Ph.D. in Chemical Physics and an MBA. As it was a golden time for physicists in finance, she quickly found a job on Wall Street. This decision sealed our fate. She had to move to New York to become a fixed income trader. So, I decided to give her a chance, take a leave from my tenured professorship, and move to New York, too. The excitement and pull of quantitative finance were so intense that, after extending my leave of absence twice, I finally decided not to return to the university and stayed in investment banking for twenty years.

**(3) What, in your opinion, is the most valuable contribution of physics to finance?**
I think that contributions of physics to finance are manifold. The principal one is the usage of the scientific method *per se*. The strong impact of the quantitative way of thinking in finance is palpable and is in sharp contrast with economics, particularly macroeconomics, which, for all practical intents and purposes, is "not even wrong." At best, it is useless, and at worst harmful, see Lipton (2016).

**(4) Which skills of your physics years did you find most useful for working in finance?**
I think that my ability to solve hyperbolic, elliptic, and, especially, parabolic equations, both analytically and numerically, helped me a lot. Also, the epistemological approach that I learned while working as an applied mathematician and physicist proved to be very valuable. In essence, finance boils down to manipulations with random variables, which is, in no small degree, what physics does, so skills acquired in my previous life, are naturally instrumental in the current setting.

**(5) Is there a specific achievement in finance of which you are most proud? If so, can you relate it to your physics background, or does it relate more to the post-physics skills you developed?**
I am very proud of being a coinventor of the local-stochastic volatility model, as well as a very potent Lewis-Lipton formula for pricing options in the stochastic volatility framework, see Lipton (2002). I think that my work in fluid dynamics and plasma physics helped me a lot in being able to formulate and solve the corresponding problems.

**(6) You have observed and participated in the organization structure of a scientific organization (i.e., a research lab or a university) as well as that of a financial institution. What differences and similarities do you see in the people and organizations and, specifically, in the people who reach leadership positions? For example, is the old dictum - "there is no democracy of people in science, but there is a democracy of ideas" - applicable to finance?**

I was able to carefully observe operations of a university as a full tenured professor and of a quant organization as co-head of the Global Quantitative Group at BofA, which at the time was one of the largest on Wall Street. I have to say that universities are more democratic, at least as far as tenured faculty is concerned. For a while, "a democracy of academic ideas" was in danger, not least because publishing of critical scientific papers became very hard, mostly due to deficiencies of the refereeing process. However, because several web depositories, such as SSRN and arXiv (where I am a moderator), are readily available, it is currently much simpler to publish a fundamental idea, than it has ever been. Of course, publishing garbage is effortless, as well. In banking, a democracy of ideas strongly depends on a particular institution. If people cannot push their ideas through, they tend to vote with their feet and move to other institutions for recognition and appropriate compensation. In this regard, finance is much more dynamic than academia.

**(8) Have you ever observed the benefits that the physics world has received from finance?**
Yes, but there are not that many, as far as I can see. True, Louis Bachelier described the Brownian motion before Einstein, but physicists completely ignored his work. More recently, some of the work on pricing American puts found applications to Stefan problem, while my work on the efficient usage of the method of heat potentials found several applications in physics, including the integrate-and-fire neuron excitation model, see Lipton (2020).

**(9) Is the value of physics to finance less than what it once was? Or do you expect the contributions of physics to finance to continue?**
I think that, at present, the value of physics is much less than it used to be due to the prevalence of data science, artificial intelligence, and similar approaches to quantitative finance. While undeniably useful, these approaches have to be strongly fortified by the scientific method. So, at some point, finance would have to rely on physics again.

**(10) Much has changed in finance and all other careers since you left physics. If you were facing that choice again now as a young physicist, would you choose finance as it is today? Or an alternative career such as coding for blockchain, cryptocurrencies, AI / ML, social media, cybersecurity, etc.? Or would you stay with physics?**

I would either stay with physics or choose an alternative career working with distributed ledger, artificial intelligence, and related concepts. After all, this is what I am doing at present.

**(11) Will finance continue to hire physicists? Or will this trend weaken or expire altogether?**

I think that finance, mainly, the buy-side, will always continue to hire physicists, but the trend is going to slow. Thankfully, right now, there many more areas where people with a physics background can find gainful employment, such as computer engineering, fintech, biotech, and others.

**(12) In terms of prestige and recognition, are positions in physics and finance similar? When you moved to finance, was it a time of frustration or excitement?**

I don't think so - it seems that a great physicist has a greater prestige than an excellent quant. Still, when I moved to finance, I was very excited and continue to be at present.

# Bibliography

Hyer T., A. Lipton-Lifschitz, and D. Pugachevsky. 1997. "Passport to success." *Risk Magazine* 10(9): 127-131.

Lebovitz, N., and A. Lifschitz. 1996. "New global instabilities of the Riemann ellipsoids." *The Astrophysical Journal* 458: 699-713.

Lifschitz, A. 1989. *Magnetohydrodynamics and Spectral Theory.* Kluwer Academic Publishers, Dordrecht, xii+446 pp.

Lifschitz, A. and E. Hameiri. 1992. "Local stability conditions in fluid dynamics." *The Physics of Fluids A* 34: 2644-2651.

Lipton, A. 1999. "Predictability and unpredictability in financial markets." *Physica D* 133: 321-347.

Lipton, A. 2001. *Mathematical Methods for Foreign Exchange: A Financial Engineer's Approach*. World Scientific, Singapore, 2001, xxii+676 pp.

Lipton, A. 2002. "The vol smile problem." *Risk Magazine* 15(2): 61-65.

Lipton, A. 2016. "Macroeconomic theories: not even wrong." *Risk Magazine* 29(9).

Lipton, A. 2018. "Blockchains and distributed ledgers in retrospective and perspective." *The Journal of Risk Finance* 19(1): 4-25.

Lipton, A. 2020. "Old problems, classical methods, new solutions." To appear.

Lipton, A., U. Pesavento, and M. Sotiropoulos. 2014. "Trading strategies via book imbalance." *Risk Magazine* 27(4): 70-75.

Lipton, A., and A. Sepp. 2008. "Stochastic volatility models and Kelvin waves." *J. Phys. A: Math. Theor.* 41: 344012 (23pp).